\documentclass[12pt]{article}

\usepackage[utf8]{inputenc} 
\usepackage[T1]{fontenc}    
\usepackage{hyperref}       
\usepackage{url}            
\usepackage{booktabs}       
\usepackage{amsfonts}       
\usepackage{nicefrac}       
\usepackage{microtype}      
\usepackage{graphicx}
\usepackage{natbib}
\usepackage{authblk}
\setcitestyle{authoryear,round,semicolon}

\title{Can Single Neurons Solve MNIST? The Computational Power of Biological Dendritic Trees
}

\author[1]{Ilenna Simone Jones}
\author[2]{Konrad Kording}
\affil[1]{Department of Neuroscience, University of Pennsylvania}
\affil[2]{Departments of Neuroscience and Bioengineering, University of Pennsylvania}

\begin{document}

\maketitle

\begin{abstract}
\indent
Physiological experiments have highlighted how the dendrites of biological neurons can nonlinearly process distributed synaptic inputs. This is in stark contrast to units in artificial neural networks that are generally linear apart from an output nonlinearity. If dendritic trees can be nonlinear, biological neurons may have far more computational power than their artificial counterparts. Here we use a simple model where the dendrite is implemented as a sequence of thresholded linear units. We find that such dendrites can readily solve machine learning problems, such as MNIST or CIFAR-10, and that they benefit from having the same input onto several branches of the dendritic tree. This dendrite model is a special case of sparse network. This work suggests that popular neuron models may severely underestimate the computational power enabled by the biological fact of nonlinear dendrites and multiple synapses per pair of neurons. The next generation of artificial neural networks may significantly benefit from these biologically inspired dendritic architectures.
\end{abstract}

\section{Introduction}
\subsection{Dendritic Nonlinearities}

Though the role of biological neurons as the mediators of sensory integration and behavioral output is clear, the computations performed within neurons has been a point of investigation for decades \citep{McCulloch1943, Hodgkin1952, FitzHugh1961, Poirazi2003, Mel2016}. For example, the McCulloch and Pitts (M\&P) neuron model is based on an approximation that a neuron linearly sums its input and maps this through a nonlinear threshold function, allowing it to carry out a selection of logic-gate-like functions, which can be expanded to create logic-based circuits \citep{McCulloch1943}. The M\&P neuron also sets the foundation for modern day neurons in artificial neural networks (ANNs), where each neuron in the network linearly sums its input and maps this through a nonlinear activation function \citep{Goodfellow2016, Lecun2015}. ANNs, made up of often millions of these neurons, are demonstrably powerful algorithms that can be trained to solve complex problems, from reinforcement learning to natural language processing to computer vision \citep{Lecun2015, Krizhevsky2006, Mnih2015, Devlin2018, Huval2015}. However, M\&P neurons and neurons of ANNs are point-neuron models that rely on linear sums of their inputs, whereas the observed physiology of biological neurons shows that dendrites impose nonlinearities on their synaptic inputs before summation at the soma \citep{London2005, Poirazi2003a, Antic2010, Agmon-Snir1998}. This indicates that M\&P and ANN neurons may radically underestimate what individual neurons can do.

Although many models of single neuron activity use linear point neurons \citep{Ujfalussy2015}, it is known that dendritic nonlinearities are responsible for a variety of neuronal dynamics and can be used to mechanistically explain the roles of biological neurons in a variety of behaviorally significant circuits \citep{London2005, Agmon-Snir1998, Barlow1965}. For example, passive properties of dendrites lead to attenuation of current along the dendrite, allowing for low-pass filtering of inputs \citep{London2005, Rall1959}. Active properties of dendrites allow for synaptic clustering to result in super-linear summation of voltage inputs upon reaching the soma \citep{Antic2010, Schiller2000, Branco2011}. These properties allow for important functions such as auditory coincidence detection and even logical operations within dendrites \citep{Mel2016, London2005, Agmon-Snir1998, Koch1983}. To fully explore the scope of biological neuron function it is then important to model more sophisticated computations within dendritic trees.

Models for individual  neurons with meaningful dendrites have been proposed to better understand neuron computation \citep{Mel2016, Gerstner2009}. Biologically detailed approaches, such as employing the multi-compartmental biophysical model \citep{Hines1997}, have been fitted to empirical data in order to study dendritic dynamics such as backpropagating action potentials and nonlinear calcium spikes \citep{London2005, Hay2011, Wilson2016}. Poirazi et al.~\citep{Poirazi2003} pioneered a more abstracted approach of modelling single neurons that isolates the impacts of including dendritic sigmoidal nonlinearities on predicting neural firing rates produced by dendrite-complete biophysical models. This novel approach used a sparsely connected two-layer ANN whose structure is analogous to that of a dendritic tree, showing it is possible to model individual neurons with ANNs.

\subsection{Repetition of synaptic inputs}

While the morphology of a dendritic tree is key to modelling its computational capabilities \citep{Mel2016, London2005, Mel1993, segev2006, Wilson2016}, it may also be important to consider the role of repeated synaptic inputs to the same postsynaptic neuron. Complex computation in ANNs depends on dense connection, which repeats inputs to each node in each layer \citep{Lecun2015}. Empirically, electron microscopy studies have shown that a presynaptic axon synapses approximately 4 times per postsynaptic neuron \citep{Kincaid1998}. Also, these studies show evidence of a certain kind of repeated synapses called multi-synaptic boutons (MSBs) \citep{Jones1997}. MSBs have shown to occur 11.5\% of the time in rats living in enriched environments \citep{Jones1997}. Additionally, it has been shown that an \emph{in vitro} long-term potentiation (LTP) induction protocol can also increase the number of MSBs of the same dendrite 6-fold \citep{Jones1997}. LTP, involved in learning and memory \citep{Bliss1973, Stuchlik2014}, can then lead to the replication of synapses between two neurons. This suggests that repeated synapses may be important for changing the computations a single neuron can do.

\subsection{Contribution}

By training and testing ANNs on complex tasks, the field of machine learning gains computational clarity \citep{Goodfellow2016, Lecun2015}. At the moment the field of neuroscience does not have this kind of in-depth computational clarity with individual, dendrite-complete neurons, despite the fact that we can describe the different behaviorally significant functions individual neurons are able to fulfill \citep{London2005, Agmon-Snir1998, Barlow1965, Gidon2020}. If we are to consider a neuron as an input/output device with a binary tree as its dendritic tree, we may be able to test its ability to learn to perform complex tasks and gain insight on how dendritic trees may impact the computation of a defined task.

Here we design a trainable, dendrite-complete neuron model in order to test its performance on  binary classification tasks taken from the field of machine learning. The model comprises a sparse ANN: a binary tree in which each nonlinear unit receives only 2 inputs. The nonlinearities and structural constraints of this ANN can be compared to a linear point-neuron model, allowing us to test the impacts of nonlinearities in a dendrite-like tree. The model also allows us to test the impact of repeated inputs on task performance. We found that our binary tree model, representing a \emph{single biological neuron}, performs better than a comparable linear classifier. Furthermore, when repeated inputs are incorporated into our model, it approximately matches the performance of a comparable 2-layer fully connected ANN. These results demonstrate that complex tasks, for which it has been assumed that an ensemble of multiple relatively simple neuron models are required, can in fact be computed by a singular, dendrite-complete neuron model.

\section{Results}

\begin{figure}[t]
    \centering
    \includegraphics[width=0.9\linewidth]{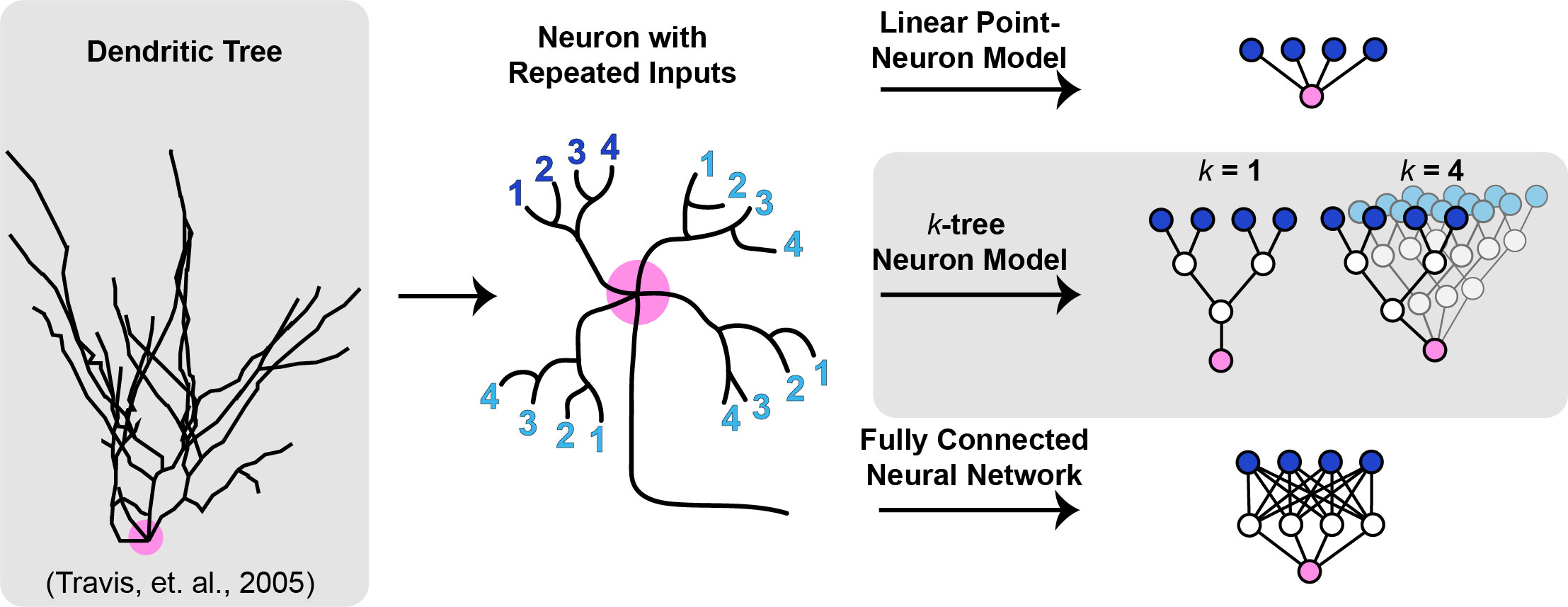}
    \caption{Novel ANN neuron model with repeated inputs. Left: Traced morphology of the dendrite of a human BA18 (occipital) cell \citep{travis2005regional}. Soma location is marked in pink. Middle: A representation of a hypothetical neuron. Inputs in dark blue at the terminal ends of one subtree are repeated in light blue in 3 other subtrees.  Upper right: Representation of a M\&P-like linear point neuron model. Middle right: $k$-tree neuron model, where $k$ = number of subtrees. Each input and hidden node has leaky ReLU activation function, and the output node has a sigmoid activation function. Bottom right: Representation of a 2-layer fully connected neural network (FCNN). Each input and hidden node has leaky ReLU activation function and the output node has a sigmoid activation function. }
    \label{fig:methods}
\end{figure}

One of the classical questions in neuroscience is how dendrite structure and the various synaptic inputs to the dendritic tree affect computation \citep{London2005, Mel2016, Rall1959}. Traditional neuron models are designed to best match observed neural dynamics \citep{Poirazi2003, Gerstner2009, Brette2011, Gouwens2018, Hay2011, Ahrens2006}, however, with exceptions \citep{Poirazi2003, Ujfalussy2015, Gidon2020, Zador1992, Zador1996, Legenstein2011}, the impacts of nonlinearities and, especially, the impacts of repeated inputs on the computational capabilities of neurons have yet to be quantified in the way we suggest. The computational abilities of ANNs can be judged by their performance on various complex tasks \citep{Goodfellow2016, Lecun2015}. Following this lead, we imposed dendritic binary tree structural constraints (Figure \ref{fig:methods}) on a trainable nonlinear ANN, resulting in a special case of sparsely connected ANN. We call this a 1-tree because it is similar to the structure of a single soma-connected subtree of a dendritic tree. (Figure \ref{fig:methods}) By repeating this subtree structure multiple times and feeding each the exact same input, we create what we call a $k$-tree, where $k$ is the number of repeated trees connected to a soma node. By using a trainable $k$-tree that has a biological structure constraint and repeated inputs, we can quantitatively judge the computational performance of this neuron model on performing complex tasks.

\begin{table}[t]
\small
\begin{center}
\begin{tabular}{@{}r|rr|rr|rr@{}}
\toprule
& \multicolumn{2}{c}{\textbf{256 inputs}}                & \multicolumn{2}{c}{\textbf{1024 inputs}}                  & \multicolumn{2}{c}{\textbf{3072 inputs}}                  \\ \hline
\textit{k} & k-tree & FCNN       & k-tree    & FCNN       & k-tree    & FCNN       \\ \hline
1          & 511    & 514     & 2,047     & 2050     & 6,143     & 6,146    \\
2          & 1,022  & 1,028    & 4,094     & 4,100     & 12,286    & 12,292     \\
4          & 2,044  & 2,056     & 8,188     & 8,200    & 24,572    & 24,584    \\
8          & 4,088  & 4,112     & 16,376    & 16,400     & 49,144    & 49,168     \\
16         & 8,176  & 8,224    & 32,752    & 32,800     & 98,288    & 98,336     \\
32         & 16,352 & 16,448    & 65,504    & 65,600     & 196,576   & 196,336     \\ \hline
\bottomrule
\end{tabular}
\vspace{5pt}
\caption{ANN Parameter Size Comparison. Fully connected neural network (FCNN) architectures are matched in parameter size to the k-tree architectures.}
\label{tab-param}
\end{center}
\end{table}

\begin{figure}[ht]
    \centering
    \includegraphics[width=0.60\linewidth]{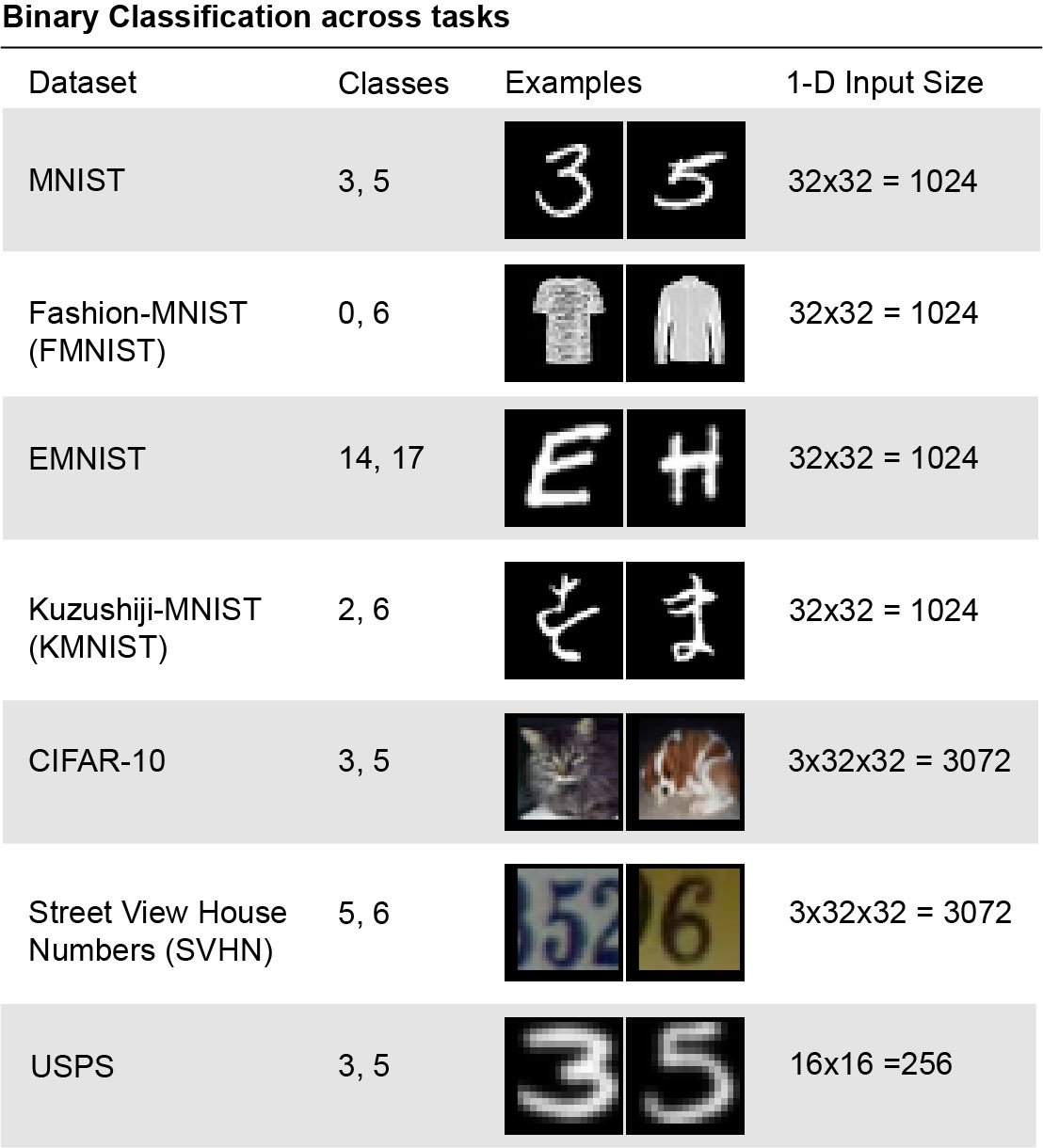}
    \caption{Classification task datasets. We considered seven machine learning datasets of varying content and size, each with ten classes. For each dataset, two of the classes were chosen by selecting the least linearly separable pair using a linear discriminant analysis (LDA) linear classifier. Each image was vectorized in order to be compatibly presented to each model. }
    \label{fig:datasets}
\end{figure}

\begin{table}[ht]
\small
\begin{center}
\begin{tabular}{r|cccc}
\toprule
& \textbf{MNIST} & \textbf{FMNIST} & \textbf{EMNIST} & \textbf{KMNIST} \\ \hline

$1$-tree & 0.9220 $\pm$   0.0179   & 0.7900 $\pm$ 0.0202  & 0.8524 $\pm$ 0.1520   & 0.8035 $\pm$ 0.0488   \\
$32$-tree & 0.9635 $\pm$   0.0043   & 0.8300 $\pm$ 0.0063   & 0.9851 $\pm$ 0.0029     & 0.8791 $\pm$ 0.0113   \\
LDA     & 0.8753 $\pm$   0.0120   & 0.6750 $\pm$ 0.0108   & 0.5821 $\pm$ 0.0180   & 0.6790 $\pm$ 0.0164  \\
1-FCNN  & 0.9546 +-   0.0053   & 0.8262 +- 0.00063     & 0.9779 +- 0.0046   & 0.8674 +- 0.0188    \\
32-FCNN     & 0.9696 $\pm$   0.0053  & 0.8290 $\pm$ 0.0075  & 0.9846 $\pm$ 0.0034   & 0.9088 $\pm$ 0.0080 \\ \hline
$1$-tree vs LDA  & $p <$ 0.0001     & $p <$ 0.0001 & $p <$ 0.0001   & $p <$ 0.0001 \\
1-tree vs 1-FCNN & $p =$ 0.0001    & $p =$ 0.0001         & $p =$ 0.0235       & $p =$ 0.0018          \\
$32$-tree vs 32-FCNN & $p =$ 0.0156 & $p =$ 0.7516 & $p =$ 0.7357   &  $p <$ 0.0001 \\  \bottomrule
\end{tabular}
\end{center}
\vspace{2pt}
\begin{center}
\begin{tabular}{r|ccc}
              \toprule & \textbf{CIFAR10} & \textbf{SVHN} & \textbf{USPS}  \\ \hline
$1$-tree        & 0.5605 $\pm$   0.0140                   & 0.5811 $\pm$ 0.0412                    & 0.8221 $\pm$ 0.0465\\
$32$-tree        & 0.5784 $\pm$ 0.0111                     & 0.6036 $\pm$ 0.0661                    & 0.8981 $\pm$ 0.0080  \\
LDA             & 0.5254 $\pm$   0.0069                   & 0.5186 $\pm$ 0.0102                    & 0.8362 $\pm$ 0.0306 \\
1-FCNN & 0.5592 +-   0.0148                   & 0.6117 +- 0.0844                    & 0.8971 +- 0.0199 \\
32-FCNN            & 0.5654 $\pm$   0.0104                   & 0.7794 $\pm$ 0.0301                    & 0.9067 $\pm$ 0.0169                                      \\ \hline
$1$-tree vs LDA   & $p <$ 0.0001  &  $p =$ 0.0005     & $p =$ 0.4897 \\
1-tree vs 1-FCNN &  $p =$ 0.8736     & $p =$ 0.4024       & $p =$ 0.0012 \\
$32$-tree vs 32-FCNN & $p =$ 0.0344 & $p <$ 0.0001     & $p =$ 0.2031 \\
\bottomrule
\end{tabular}
\end{center}
\vspace{5pt}
\caption{$k$-tree Mean Performance Comparison to FCNN and LDA. Performance accuracy is listed as mean $\pm$ standard error for a set of 10 trials. $p$-Values calculated using student's t-test. LDA and FCNN are used as lower and upper bounds that the $k$-tree is compared to.}
\label{tab-ktree}
\end{table}

Neurons, arguably, produce binary outputs (presence or absence of an action potential) \citep{Hodgkin1952}. Therefore, to fairly judge an individual neuron model’s performance on a complex task, we will use a binary classification task. The complexity in the tasks can come from high-dimensional vector inputs from images taken from classic computer vision datasets used in the field of machine learning (Figure \ref{fig:datasets}).

As controls for performance comparison, we used a linear discriminant analysis (LDA) linear classifier to approximate the performance of a linear point neuron model, and a fully connected neural network (FCNN) that is comparable in size to the $k$-tree. The linear classifier model is relatively simple compared to the more parameter-complex $k$-tree and FCNN, and we expect it to be able to learn fewer functions \citep{Dreiseitl2002}; therefore, its performance sets an expected lower-bound. The FCNN is densely connected and consists of 2-layers. With its nonlinearities, we expect it to learn to express a greater variety of functions, therefore its performance sets an expected upper bound. To compare the two ANNs, let us say that $n$ is the number of pixel inputs to each classifier, determining the number of parameters, $P$, needed in each network, and $h$ is the number of nodes in the hidden layer of the FCNN. Based on the constraints of each network, the FCNN will then have $P = h(n + 1)$ and the $k$-tree will have $P = k(2n - 1)$. To match the number of parameters of the FCNN to that of the $k$-tree, we assert that $h = 2k$. (Table \ref{tab-param}).

\subsection{Nonlinear tree neuron model performs better than a linear classifier}

\begin{figure}[ht]
    \centering
    \includegraphics[width=0.9\linewidth]{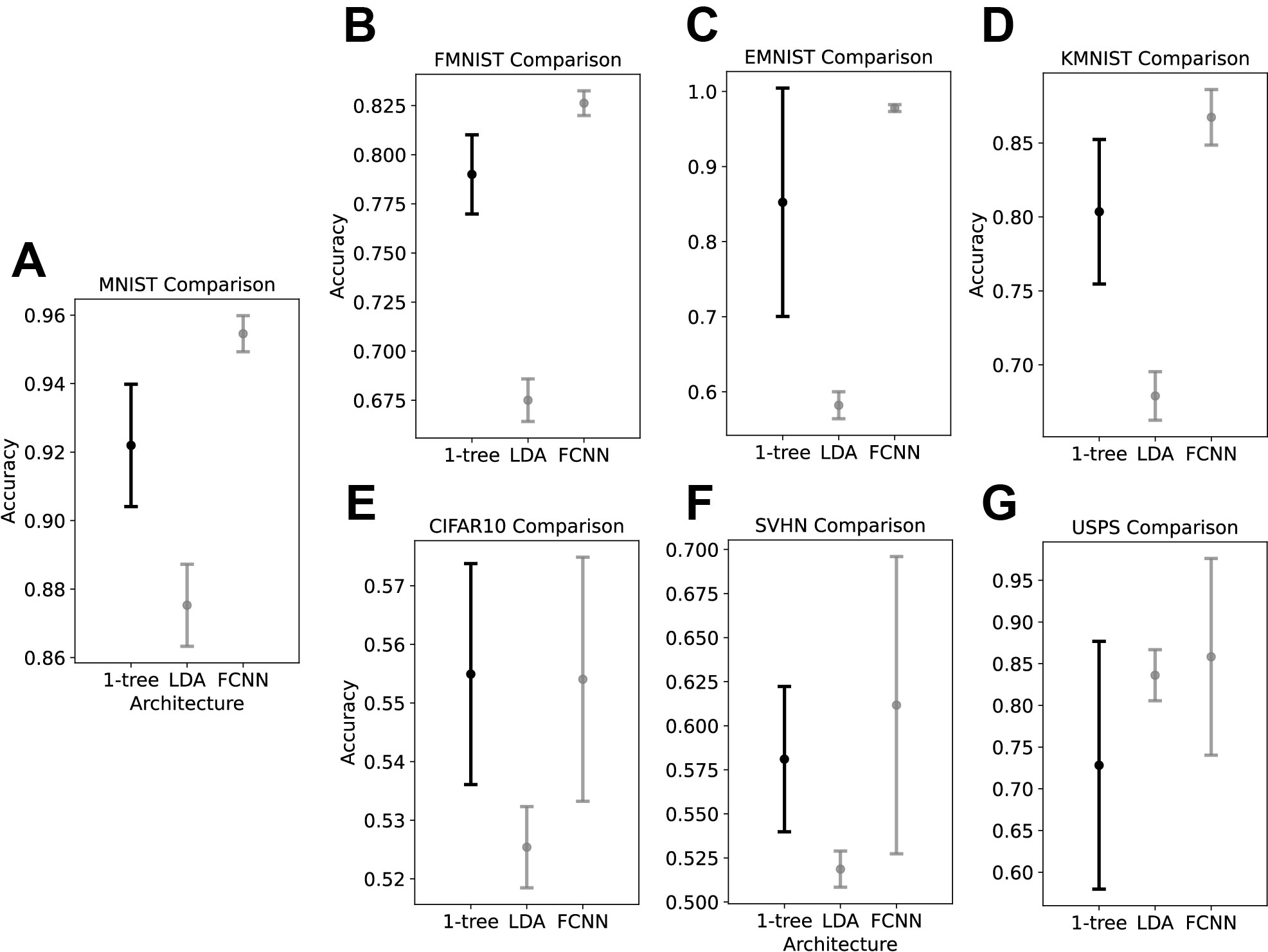}
    \caption{Performance of 1-tree compared to linear classifier and FCNN. 1-tree performance is compared to that of a lower bound, LDA, and an upper bound, FCNN. For most tasks, the 1-tree performs better than LDA, and FCNN performs better than the 1-tree.}
    \label{fig:1tree}
\end{figure}

Classical models of neurons have been of linear point neurons that do not take into consideration dendritic nonlinearities \citep{McCulloch1943, Hodgkin1952, FitzHugh1961}. By considering dendritic nonlinearity and structure, we designed a new neuron model: a nonlinear ANN with the structural constraints of a dendritic tree called a 1-tree. We then compared the performance of this new model against a proxy for a point neuron, an LDA linear classifier. Focusing on one simple image classification task of a binary dataset of handwritten numbers, MNIST, we compare the computational performance of the 1-tree and the linear classifier and the nonlinear, structurally dendritic 1-tree. Significantly, the performance of the 1-tree is greater than that of the linear classifier with $p < $ 0.0001 (Figure \ref{fig:1tree}A, Table \ref{tab-ktree}). Repeating this test with 6 more datasets, we find that for most of these datasets, regardless of the size of the input dimensionality in each dataset, the 1-tree performs consistently above the linear classifier (FMNIST, EMNIST, KMNIST, CIFAR10: $p < $ 0.0001; SVHN: $p = $ 0.0005) (Figure \ref{fig:1tree}B-F, Table \ref{tab-ktree}). Exceptionally, the performance of the 1-tree on the USPS dataset had no significant difference of performance compared tothe linear classifier. The USPS dataset has the smallest input dimensionality (256 pixels) and leads to a 1-tree with the fewest parameters (P = 511, Table \ref{tab-param}). It could be the network was not complex enough to perform better than the linear classifier. Barring this exception, not only do dendrites have nonlinear properties, nonlinearities in a dendrite-like neuron model generally improves its computational performance compared to that of a linear classifier.

For comparison to the 1-tree, we tested a 2-layer fully connected neural network (FCNN) matched in parameter size to the 1-tree. In the MNIST task, the FCNN performed significantly better than the 1-tree with a $p = $ 0.0001 (Figure \ref{fig:1tree}A, Table \ref{tab-ktree}). We then tested the 6 additional datasets, resulting in differently sized 1-trees and FCNNs due to differences in input sizes. The similarly sized FMNIST, EMNIST, and KMNIST dataset networks maintained the significant difference between the 1-tree and FCNN (Figure \ref{fig:1tree}B-D, Table \ref{tab-ktree}). The USPS dataset also maintained a significant difference (Figure \ref{fig:1tree}G). The CIFAR10 and SVHN datasets did not have a significant difference in performance (Figure \ref{fig:1tree}E-F, Table \ref{tab-ktree}). The high variance in the FCNN performance for CIFAR10 and SVHN (Figure \ref{fig:1tree}E-F, Table \ref{tab-ktree}) may be due to the FCNN’s failure to train in some trials, resulting in performances close to 50\%. For most tasks we tried, the FCNN performed much better than the 1-tree.

\subsection{Repeating inputs to tree model increases performance comparable to FCNN with a small fraction of the parameters}
\begin{figure}
    \centering
    \includegraphics[width=0.9\linewidth]{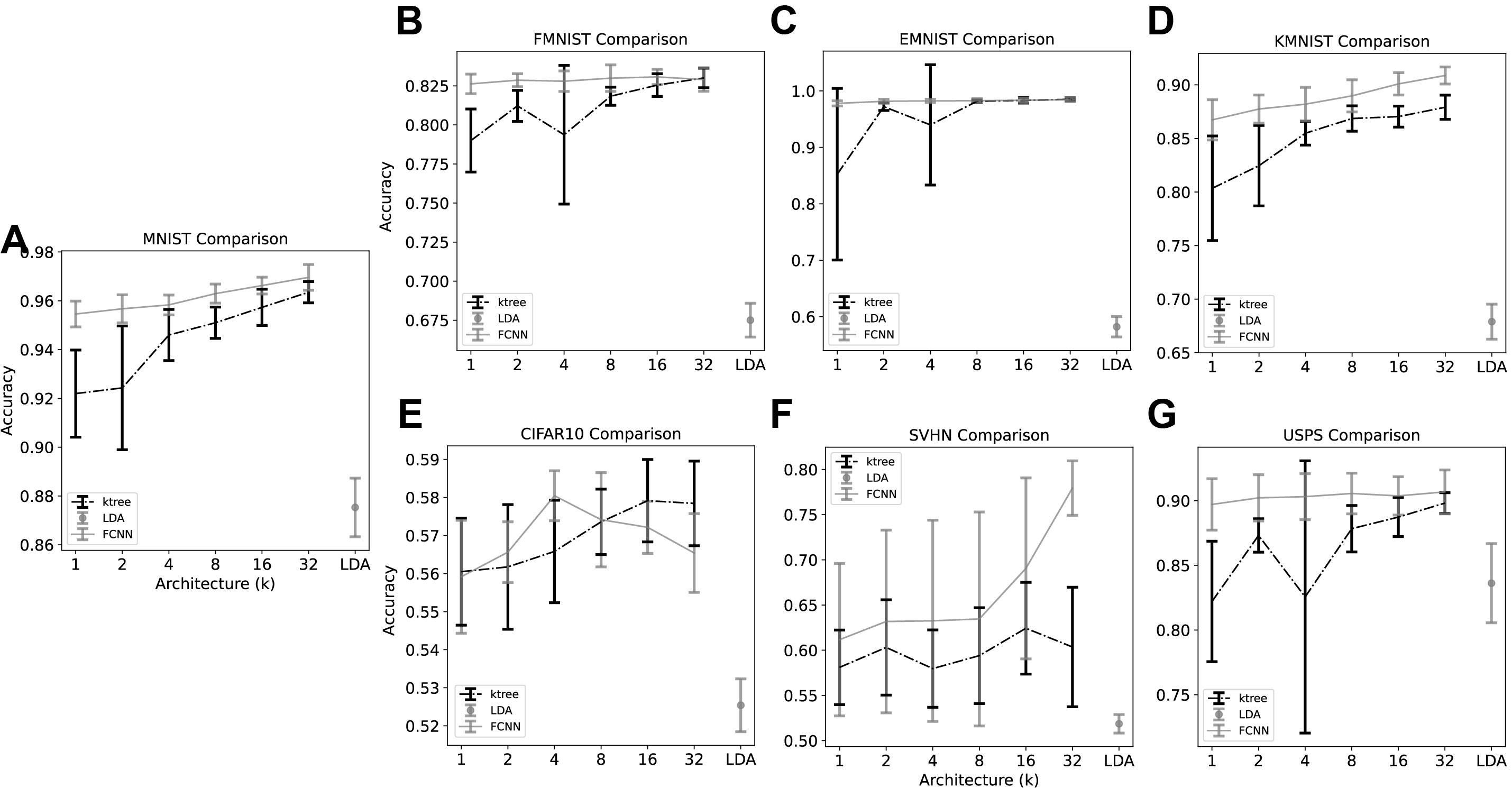}
    \caption{Performance of $k$-tree compared to a linear classifier and FCNN. $k$-tree performance is compared to that of a lower bound, LDA, and an upper bound, FCNN. The $k$ is doubled 5 times, resulting in tests of $k$ = {1, 2, 4, 8, 16, 32}. In all cases, as $k$ (the number of repeated dendritic subtrees) increases, so does the performance accuracy of the $k$-tree, approaching the upper bound.}
    \label{fig:ktree}
\end{figure}
The computational impact of repeated inputs to a dendritic tree is not clear, however studies have shown increased repetition of inputs as a result of plasticity events \citep{Toni1999}, which has implications for learning and memory. By modifying the 1-tree by repeating the tree structure and input to the model $k$ times, we can then achieve a $k$-tree neuron model (Figure \ref{fig:methods}). This can be a proxy for seeing how repeated inputs might impact computational performance on various binary image classification tasks. Returning to the MNIST dataset, we tested $k$ = {1, 2, 4, 8, 16, 32} and observed how increasing $k$ can gradually improve performance. For example, compare the performance of a 1-tree to that of a  32-tree in the MNIST binary classification task ($p =$ 0.0000) (Figure \ref{fig:ktree}A). Remarkably, the performance of the 32-tree (96.35 $\pm$ 0.43\%) is very close to the performance of the FCNN (96.96 $\pm$ 0.53\%), yet still different with $p = $ 0.0156 (Figure \ref{fig:ktree}A, Table \ref{tab-ktree}). Increasing the number of repeats to a $k$-tree neuron model improves its performance on the MNIST binary classification task, nearly meeting the performance of a comparable FCNN.

In order to see if this result generalizes, we tested the $k$-tree on 6 additional binary image classification datasets. All tasks see an increase in performance as the number of subtrees in the $k$-tree increases up to $k$ = 32 (Figure \ref{fig:ktree}B-G). The 32-tree has meets the performance of the FCNN in the FMNIST ($p =$ 0.7516), EMNIST ($p =$ 0.7357), and USPS ($p = $ 0.2031) tasks (Figure \ref{fig:ktree}B,C,G, Table \ref{tab-ktree}). For the CIFAR10 dataset, FCNN performance peaks at $k =$ 4, then decreases, resulting in the 32-tree surpassing the 32-FCNN (Figure \ref{fig:ktree}E, Table \ref{tab-ktree}). We can then say that increasing the number of repeats to a $k$-tree neuron model improves its computational performance in all tasks such that it approaches the performance of a comparable FCNN.

\section{Methods}

\subsection{Computational Tasks}
Knowing that the output of a neuron is binary (presence or absence of an action potential), we chose to train our neuron model on a binary classification task. Using standard, high-dimensional, computer vision datasets, we used linear discriminant analysis (LDA) linear classifier to determine which 2 classes within each dataset were least linearly separable through training the LDA linear classifier and testing it on pairs of classes (Figure \ref{fig:datasets}). We used MNIST \citep{Lecun1998}, Fashion-MNIST \citep{Xiao2017}, EMNIST \citep{Cohen2017}, Kuzushiji-MNIST \citep{Clanuwat2018}, CIFAR-10 \citep{Krizhevsky2009}, Street View House Numbers (SVHN) \citep{Goodfellow2014}, and USPS \citep{Hastie2001} datasets.

\subsection{Controls}
The controls we use are the LDA linear classifier and a fully connected neural network (FCNN). The linear classifier sets a baseline performance for linear separability of each of the two classes per dataset, in addition to acting as a proxy for a linear point neuron model. The 2-layer FCNN is a comparable reference to see if $k$-tree performance meets or exceeds that of a densely connected network. The hidden layer of the FCNN is equal to twice the number of trees ($2k$) in the k-tree it is compared to and its output layer has 1 node.

\subsection{Data Preprocessing}
We used datasets from the torchvision (version 0.5.0) python package. We then padded the 28 by 28 resolution images with zeros so that they were 32 x 32, and flattened the images to 1-D vectors. We then split the shuffled training set into training and validation sets (for MNIST, the ratio was 1:5 so as to let the validation set size match the test set), Then we split the resultant shuffled training set and validation set into 10 independent subsets. Each set was used for a different cross-validation trial.

\subsection{Model Architecture}
Using PyTorch (version 1.4.0), we designed the $k$-tree model architecture to be a feed forward neural network with sparse binary-tree connections. The weight matrices, which were dense tensors, of each layer were sparsified such that each node receives 2 inputs and produces 1 output. For example, the 1024 pixel-size images were fed to a 1-tree with 10 layers: the input layer is 1024 by 512, the 2nd layer 512 by 256 etc. until the penultimate layer is reached with dimensions 2 by 1. The final layer is $k$ by 1 where $k$ is the number of subtrees in the $k$-tree; in this case it would be 1 by 1. In the special case of the 3072 pixel size images, inputs were fed into a 1-tree with 11 layers, the input layer is 3072 by 1024, the 2nd layer is 1024 by 512, etc.

To account for the sparsification, we altered the initialization of the weight matrices: we used standard ``Kaiming normal'' initialization with a gain of 1/density of sparsified dense tensor weight matrices. We also created a ``freeze mask'' that recorded which weights were set to 0 in order to freeze those weights during training later. For the forward step, we used leaky ReLU with a slope of 0.01 for nodes between layers, and sigmoid nonlinearity at the final output node which kept output values between 0 and 1.

\subsection{Model Training}
The model, inputs, and labels were loaded onto a Nvidia GeForce 1080 GPU using CUDA version 10.1. The batch size was 256. Early stopping was used such that after 60 epochs where no decrease in the loss is observed, training is stopped. Loss was calculated using binary cross entropy loss. We used an Adam optimizer with a learning rate of 0.001. Within the training loop immediately after the backward step and before updating the weights using the gradients, we zeroed out the gradients indicated by the freeze mask so as to keep the model sparsely connected. Each train-test loop was run for 10 trials with a different training subset each trial and the same test set every trial. Trial averages and standard deviation were then calculated and p-values were calculated using student t-test.

\section{Discussion}
\raggedbottom

Here we quantify the potential computational capabilities of an abstracted neuron model with dendritic features and repeated inputs. We designed a trainable neuron model: a sparse ANN with binary dendritic tree constraints made up of nonlinear nodes (Figure \ref{fig:methods}). The tree that resulted from this constraint was repeated $k$ times with identical inputs in order to explore the impacts of repeated inputs. We judged the model by determining its performance on 7 high-dimensional binary image classification tasks (Figure \ref{fig:datasets}), and compared its performance to a linear classifier, a lower bound, and a comparable FCNN, an upper bound. The 1-tree, with its nonlinear nodes and dendritic structure constraint, performed better than the linear classifier in almost all tasks (Figure \ref{fig:1tree}). When we increased $k$ of the $k$-tree from $k$ = 1 to $k$ = 32, we saw a consistent increase in $k$-tree performance across all tasks (Figure \ref{fig:ktree}). In the case of the MNISt task, the performance of the 32-tree was close to the comparable FCNN performance. Surprisingly, the 32-tree in the FMNIST, EMNIST, and USPS tasks met that of the comparable FCNN. These findings emphasize the importance for modelers to consider both dendrites and synaptic input repetitions.

A limitation of this study is the relevance of our computational tasks. Although it is hard to know exactly what kind of input a neuron receives from its presynaptic connections, we do not believe the 1-dimensional vectorized input we provide our neuron model is biologically plausible. Ordering the pixel input to these models randomly overall decreases k-tree performance, implying that the order of the input impacts performance (see Figure S1 in the Supplementary Material). Further investigation may be needed to explore how the ordering of the 1-D pixel input might impact performance.

The binary tree structure we chose to constrain an ANN to make the $k$-tree makes several assumptions. Each node of the tree is analogous to a compartment in a dendritic tree, and in biological dendritic trees each compartment will receive an exclusive set of inputs. Therefore, we chose not to use convolution or any kind of weight sharing in our model. In addition, the synaptic weights and inter-node weights are real-valued free parameters, however the weights analogous to inter-compartmental axial resistances \citep{Rall1959,Huys2006} could only be positive scalar values if biologically plausible. Future work to address this would be to constrain the free parameter ranges to be completely positive.

In this study we used an abstracted model to give us insights into the impacts of biological constraints and properties. After all, these kinds of optimizations are not currently doable in more realistic models of neurons. Using this model, we see how nonlinear dendrites increase a neuron model’s task performance above that of a linear classifier, which serves as a proxy for models following the point-neuron assumption. Importantly, we see how by repeating the inputs to this dendrite model we can observe a consistent increase in task performance. These findings emphasize the importance for modelers to consider both dendrites and synaptic input repetitions.

Our results may also be relevant for the field of deep learning. The $k$-trees we consider are special cases of sparse ANN, wherein there are only 2 inputs to all nodes after the first layer. These contrast with randomly-made sparse networks or pruned sparse networks \citep{Frankle2019}, because they have very severe constraints. It is then surprising that a $k$-tree could perform at the level of a comparable FCNN. We would be interested in future work comparing the performance of binary tree structures, inspired by biological dendrites, against the performance of less structured sparse ANNs with comparable edge density. 

This study tests the classification performance of a dendrite-complete neuron model and compares it to a model that follows the point-neuron assumption, highlighting the importance of considering branching dendrite structure and nonlinearities when modeling neurons. We expand this test to consider the possibility of repeated synaptic inputs in our model, showing that the model consistently performs better with more repeated inputs to additional subtrees. We also see that the sparse network neuron model we designed can reach similar performance to a comparable densely connected network. Fundamentally, this study is a foray into directly considering a neuron’s computational capability by training the model to perform complex tasks using deep learning methodology, which promises to further our insights into single neuron computation.

\section{Acknowledgements}
I would like to thank the members of the Kording Lab, specifically Roozbeh Farhoodi, Ari Benjamin, and David Rolnick for their help over the development of this project.

\section{Code}
The code for this project can be found at the following github repository: https://github.com/ilennaj/ktree

\medskip

\small
\raggedbottom
\pagebreak

\bibliography{bibliography}

\begin{thebibliography}{50}
\providecommand{\natexlab}[1]{#1}
\providecommand{\url}[1]{\texttt{#1}}
\expandafter\ifx\csname urlstyle\endcsname\relax
  \providecommand{\doi}[1]{doi: #1}\else
  \providecommand{\doi}{doi: \begingroup \urlstyle{rm}\Url}\fi

\bibitem[Agmon-Snir et~al.(1998)Agmon-Snir, Carr, and Rinzel]{Agmon-Snir1998}
H.~Agmon-Snir, C.~E. Carr, and J.~Rinzel.
\newblock {The role of dendrites in auditory coincidence detection}.
\newblock \emph{Nature}, 393:\penalty0 268--272, 1998.

\bibitem[Ahrens et~al.(2006)Ahrens, Huys, and Paninski]{Ahrens2006}
M.~B. Ahrens, Q.~J.~M. Huys, and L.~Paninski.
\newblock {Large-scale biophysical parameter estimation in single neurons via
  constrained linear regression}.
\newblock \emph{Advances in Neural Information Processing Systems}, 2006.
\newblock ISSN 10495258.
\newblock \doi{10.1007/s00439-005-0104-y}.
\newblock URL
  \url{http://citeseerx.ist.psu.edu/viewdoc/download?doi=10.1.1.184.2712{\&}rep=rep1{\&}type=pdf}.

\bibitem[Antic et~al.(2010)Antic, Zhou, Moore, Short, and Ikonomu]{Antic2010}
S.~D. Antic, W.~L. Zhou, A.~R. Moore, S.~M. Short, and K.~D. Ikonomu.
\newblock {The decade of the dendritic NMDA spike}.
\newblock \emph{Journal of Neuroscience Research}, 88\penalty0 (14):\penalty0
  2991--3001, 2010.
\newblock ISSN 03604012.
\newblock \doi{10.1002/jnr.22444}.

\bibitem[Barlow and Levick(1965)]{Barlow1965}
H.~B. Barlow and W.~R. Levick.
\newblock {The mechanism of directionally selective units in rabbit's retina.}
\newblock \emph{The Journal of Physiology}, 178\penalty0 (3):\penalty0
  477--504, 1965.
\newblock ISSN 14697793.
\newblock \doi{10.1113/jphysiol.1965.sp007638}.

\bibitem[Bliss and Lomo(1973)]{Bliss1973}
T.~V.~P. Bliss and T.~Lomo.
\newblock {Long‐lasting potentiation of synaptic transmission in the dentate
  area of the unanaesthetized rabbit following stimulation of the perforant
  path}.
\newblock \emph{The Journal of Physiology}, 232\penalty0 (2):\penalty0
  357--374, 1973.
\newblock ISSN 14697793.
\newblock \doi{10.1113/jphysiol.1973.sp010274}.

\bibitem[Branco and H{\"{a}}usser(2011)]{Branco2011}
T.~Branco and M.~H{\"{a}}usser.
\newblock {Synaptic Integration Gradients in Single Cortical Pyramidal Cell
  Dendrites}.
\newblock \emph{Neuron}, 69\penalty0 (5):\penalty0 885--892, 2011.
\newblock ISSN 08966273.
\newblock \doi{10.1016/j.neuron.2011.02.006}.

\bibitem[Brette et~al.(2011)Brette, Fontaine, Magnusson, Rossant, Platkiewicz,
  and Goodman]{Brette2011}
R.~Brette, B.~Fontaine, A.~K. Magnusson, C.~Rossant, J.~Platkiewicz, and
  D.~F.~M. Goodman.
\newblock {Fitting Neuron Models to Spike Trains}.
\newblock \emph{Frontiers in Neuroscience}, 5\penalty0 (February):\penalty0
  1--8, 2011.
\newblock \doi{10.3389/fnins.2011.00009}.

\bibitem[Clanuwat et~al.(2018)Clanuwat, Bober-Irizar, Kitamoto, Lamb, Yamamoto,
  and Ha]{Clanuwat2018}
T.~Clanuwat, M.~Bober-Irizar, A.~Kitamoto, A.~Lamb, K.~Yamamoto, and D.~Ha.
\newblock {Deep Learning for Classical Japanese Literature}.
\newblock \emph{Advances in Neural Information Processing Systems}, pages 1--8,
  2018.
\newblock \doi{10.20676/00000341}.
\newblock URL
  \url{http://arxiv.org/abs/1812.01718{\%}0Ahttp://dx.doi.org/10.20676/00000341}.

\bibitem[Cohen et~al.(2017)Cohen, Afshar, Tapson, and {Van Schaik}]{Cohen2017}
G.~Cohen, S.~Afshar, J.~Tapson, and A.~{Van Schaik}.
\newblock {EMNIST: Extending MNIST to handwritten letters}.
\newblock \emph{Proceedings of the International Joint Conference on Neural
  Networks}, 2017-May:\penalty0 2921--2926, 2017.
\newblock \doi{10.1109/IJCNN.2017.7966217}.

\bibitem[Devlin et~al.(2018)Devlin, Chang, Lee, and Toutanova]{Devlin2018}
J.~Devlin, M.-W. Chang, K.~Lee, and K.~Toutanova.
\newblock {BERT: Pre-training of Deep Bidirectional Transformers for Language
  Understanding}.
\newblock \emph{arXiv preprint}, 2018.
\newblock URL \url{http://arxiv.org/abs/1810.04805}.

\bibitem[Dreiseitl and Ohno-Machado(2002)]{Dreiseitl2002}
S.~Dreiseitl and L.~Ohno-Machado.
\newblock {Logistic regression and artificial neural network classification
  models: A methodology review}.
\newblock \emph{Journal of Biomedical Informatics}, 35\penalty0 (5-6):\penalty0
  352--359, 2002.
\newblock ISSN 15320464.
\newblock \doi{10.1016/S1532-0464(03)00034-0}.

\bibitem[FitzHugh(1961)]{FitzHugh1961}
R.~FitzHugh.
\newblock {Impulses and Physiological States in Theoretical Models of Nerve
  Membrane}.
\newblock \emph{Biophysical Journal}, 1\penalty0 (6):\penalty0 445--466, 1961.
\newblock ISSN 00063495.
\newblock \doi{10.1016/S0006-3495(61)86902-6}.
\newblock URL \url{http://dx.doi.org/10.1016/S0006-3495(61)86902-6}.

\bibitem[Frankle and Carbin(2019)]{Frankle2019}
J.~Frankle and M.~Carbin.
\newblock {The lottery ticket hypothesis: Finding sparse, trainable neural
  networks}.
\newblock \emph{7th International Conference on Learning Representations, ICLR
  2019}, pages 1--42, 2019.

\bibitem[Gerstner and Naud(2009)]{Gerstner2009}
W.~Gerstner and R.~Naud.
\newblock {How good are neuron models?}
\newblock \emph{Science}, 326\penalty0 (5951):\penalty0 379--380, 2009.
\newblock ISSN 00368075.
\newblock \doi{10.1126/science.1181936}.

\bibitem[Gidon et~al.(2020)Gidon, Zolnik, Fidzinski, Bolduan, Papoutsi,
  Poirazi, Holtkamp, Vida, and Larkum]{Gidon2020}
A.~Gidon, T.~A. Zolnik, P.~Fidzinski, F.~Bolduan, A.~Papoutsi, P.~Poirazi,
  M.~Holtkamp, I.~Vida, and M.~E. Larkum.
\newblock {Dendritic action potentials and computation in human layer 2/3
  cortical neurons.}
\newblock \emph{Science (New York, N.Y.)}, 367\penalty0 (6473):\penalty0
  83--87, 2020.
\newblock ISSN 1095-9203.
\newblock \doi{10.1126/science.aax6239}.
\newblock URL \url{http://www.ncbi.nlm.nih.gov/pubmed/31896716}.

\bibitem[Goodfellow et~al.(2016)Goodfellow, Bengio, and
  Courville]{Goodfellow2016}
I.~Goodfellow, Y.~Bengio, and A.~Courville.
\newblock \emph{{Deep Learning}}.
\newblock MIT Press, 2016.
\newblock URL \url{www.deeplearningbook.org}.

\bibitem[Goodfellow et~al.(2014)Goodfellow, Bulatov, Ibarz, Arnoud, and
  Shet]{Goodfellow2014}
I.~J. Goodfellow, Y.~Bulatov, J.~Ibarz, S.~Arnoud, and V.~Shet.
\newblock {Multi-digit number recognition from street view imagery using deep
  convolutional neural networks}.
\newblock \emph{2nd International Conference on Learning Representations, ICLR
  2014 - Conference Track Proceedings}, pages 1--13, 2014.

\bibitem[Gouwens et~al.(2018)Gouwens, Berg, Feng, Sorensen, Zeng, Hawrylycz,
  Koch, and Arkhipov]{Gouwens2018}
N.~W. Gouwens, J.~Berg, D.~Feng, S.~A. Sorensen, H.~Zeng, M.~J. Hawrylycz,
  C.~Koch, and A.~Arkhipov.
\newblock {Systematic generation of biophysically detailed models for diverse
  cortical neuron types}.
\newblock \emph{Nature Communications}, 9\penalty0 (1), 2018.
\newblock ISSN 20411723.
\newblock \doi{10.1038/s41467-017-02718-3}.
\newblock URL \url{http://dx.doi.org/10.1038/s41467-017-02718-3}.

\bibitem[Hastie et~al.(2001)Hastie, Tibshirani, and Friedman]{Hastie2001}
T.~Hastie, R.~Tibshirani, and J.~Friedman.
\newblock \emph{{The elements of statistical learning}}.
\newblock Springer-Verlag, 2001.

\bibitem[Hay et~al.(2011)Hay, Hill, Sch{\"{u}}rmann, Markram, and
  Segev]{Hay2011}
E.~Hay, S.~Hill, F.~Sch{\"{u}}rmann, H.~Markram, and I.~Segev.
\newblock {Models of neocortical layer 5b pyramidal cells capturing a wide
  range of dendritic and perisomatic active properties}.
\newblock \emph{PLoS Computational Biology}, 7\penalty0 (7), 2011.
\newblock ISSN 1553734X.
\newblock \doi{10.1371/journal.pcbi.1002107}.

\bibitem[Hines and Carnevale(1997)]{Hines1997}
M.~L. Hines and N.~T. Carnevale.
\newblock {The NEURON simulation environment.}
\newblock \emph{Neural computation}, 9\penalty0 (6):\penalty0 1179--209, 1997.
\newblock ISSN 0899-7667.
\newblock URL \url{http://www.ncbi.nlm.nih.gov/pubmed/9248061}.

\bibitem[Hodgkin and Huxley(1952)]{Hodgkin1952}
Hodgkin and Huxley.
\newblock {A quantitative description of membrane current and its application
  to conduction and excitation in nerve.}
\newblock \emph{J Physiology}, 1117:\penalty0 500--544, 1952.
\newblock ISSN 09237984.
\newblock \doi{10.1080/00062278.1939.10600645}.

\bibitem[Huval et~al.(2015)Huval, Wang, Tandon, Kiske, Song, Pazhayampallil,
  Andriluka, Rajpurkar, Migimatsu, Cheng-Yue, Mujica, Coates, and
  Ng]{Huval2015}
B.~Huval, T.~Wang, S.~Tandon, J.~Kiske, W.~Song, J.~Pazhayampallil,
  M.~Andriluka, P.~Rajpurkar, T.~Migimatsu, R.~Cheng-Yue, F.~Mujica, A.~Coates,
  and A.~Y. Ng.
\newblock {An Empirical Evaluation of Deep Learning on Highway Driving}.
\newblock \emph{arXiv preprint}, pages 1--7, 2015.
\newblock URL \url{http://arxiv.org/abs/1504.01716}.

\bibitem[Huys et~al.(2006)Huys, Ahrens, and Paninski]{Huys2006}
Q.~J.~M. Huys, M.~B. Ahrens, and L.~Paninski.
\newblock {Efficient Estimation of Detailed Single-Neuron Models}.
\newblock \emph{Journal of Neurophysiology}, 96\penalty0 (2):\penalty0
  872--890, 2006.
\newblock ISSN 0022-3077.
\newblock \doi{10.1152/jn.00079.2006}.

\bibitem[Jones et~al.(1997)Jones, Klintsova, Kilman, Sirevaag, and
  Greenough]{Jones1997}
T.~A. Jones, A.~Y. Klintsova, V.~L. Kilman, A.~M. Sirevaag, and W.~T.
  Greenough.
\newblock {Induction of multiple synapses by experience in the visual cortex of
  adult rats}.
\newblock \emph{Neurobiology of Learning and Memory}, 68\penalty0 (1):\penalty0
  13--20, 1997.
\newblock ISSN 10747427.
\newblock \doi{10.1006/nlme.1997.3774}.

\bibitem[Kincaid et~al.(1998)Kincaid, Zheng, and Wilson]{Kincaid1998}
A.~E. Kincaid, T.~Zheng, and C.~J. Wilson.
\newblock {Connectivity and convergence of single corticostriatal axons}.
\newblock \emph{Journal of Neuroscience}, 18\penalty0 (12):\penalty0
  4722--4731, 1998.
\newblock ISSN 02706474.
\newblock \doi{10.1523/jneurosci.18-12-04722.1998}.

\bibitem[Koch et~al.(1983)Koch, Poggio, and Torre]{Koch1983}
C.~Koch, T.~Poggio, and V.~Torre.
\newblock {Nonlinear interactions in a dendritic tree: Localization, timing,
  and role in information processing}.
\newblock \emph{Proceedings of the National Academy of Sciences of the United
  States of America}, 80\penalty0 (May):\penalty0 2799--2802, 1983.

\bibitem[Krizhevsky(2009)]{Krizhevsky2009}
A.~Krizhevsky.
\newblock {Learning multiple layers of features from tiny images}.
\newblock \emph{ArXiv}, 2009.
\newblock ISSN 00012475.

\bibitem[Krizhevsky et~al.(2006)Krizhevsky, Sutskever, and
  Hinton]{Krizhevsky2006}
A.~Krizhevsky, I.~Sutskever, and G.~E. Hinton.
\newblock {ImageNet Classification with Deep Convolutional Neural Networks}.
\newblock \emph{Advances in Neural Information Processing Systems}, 8:\penalty0
  713--772, 2006.
\newblock \doi{10.1016/B978-008046518-0.00119-7}.

\bibitem[Lecun et~al.(1998)Lecun, Bottou, Bengio, and Haffner]{Lecun1998}
Y.~Lecun, L.~Bottou, Y.~Bengio, and P.~Haffner.
\newblock {Gradient-based learning applied to document recognition}.
\newblock \emph{proc. of the IEEE}, 1998.
\newblock URL
  \url{http://ieeexplore.ieee.org/document/726791/{\#}full-text-section}.

\bibitem[Lecun et~al.(2015)Lecun, Bengio, and Hinton]{Lecun2015}
Y.~Lecun, Y.~Bengio, and G.~Hinton.
\newblock {Deep learning}.
\newblock \emph{Nature}, 521\penalty0 (7553):\penalty0 436--444, 2015.
\newblock ISSN 14764687.
\newblock \doi{10.1038/nature14539}.

\bibitem[Legenstein and Maass(2011)]{Legenstein2011}
R.~Legenstein and W.~Maass.
\newblock {Branch-specific plasticity enables self-organization of nonlinear
  computation in single neurons}.
\newblock \emph{Journal of Neuroscience}, 31\penalty0 (30):\penalty0
  10787--10802, 2011.
\newblock ISSN 02706474.
\newblock \doi{10.1523/JNEUROSCI.5684-10.2011}.

\bibitem[London and H{\"{a}}usser(2005)]{London2005}
M.~London and M.~H{\"{a}}usser.
\newblock {Dendritic Computation}.
\newblock \emph{Annual Review of Neuroscience}, 28\penalty0 (1):\penalty0
  503--532, 2005.
\newblock ISSN 0147-006X.
\newblock \doi{10.1146/annurev.neuro.28.061604.135703}.
\newblock URL
  \url{http://www.annualreviews.org/doi/10.1146/annurev.neuro.28.061604.135703}.

\bibitem[McCulloch and Pitts(1943)]{McCulloch1943}
W.~S. McCulloch and W.~Pitts.
\newblock {A logical calculus of the ideas immanent in nervous activity}.
\newblock \emph{The Bulletin of Mathematical Biophysics}, 5\penalty0
  (4):\penalty0 115--133, 1943.
\newblock ISSN 00074985.
\newblock \doi{10.1007/BF02478259}.

\bibitem[Mel(2016)]{Mel2016}
B.~Mel.
\newblock {Toward a simplified model of an active dendritic tree}.
\newblock In G.~J. Stuart, N.~Spruston, and M.~H{\"{a}}usser, editors,
  \emph{Dendrites}. Oxford Scholarship Online, 2016.
\newblock ISBN 9780199682676.
\newblock \doi{10.1093/acprof}.

\bibitem[Mel(1993)]{Mel1993}
B.~W. Mel.
\newblock {Synaptic integration in an excitable dendritic tree.}
\newblock \emph{Journal of neurophysiology}, 70\penalty0 (3):\penalty0
  1086--101, 1993.
\newblock ISSN 0022-3077.
\newblock \doi{10.1152/jn.1993.70.3.1086}.
\newblock URL \url{http://www.ncbi.nlm.nih.gov/pubmed/8229160}.

\bibitem[Mnih et~al.(2015)Mnih, Kavukcuoglu, Silver, Rusu, Veness, Bellemare,
  Graves, Riedmiller, Fidjeland, Ostrovski, Petersen, Beattie, Sadik,
  Antonoglou, King, Kumaran, Wierstra, Legg, and Hassabis]{Mnih2015}
V.~Mnih, K.~Kavukcuoglu, D.~Silver, A.~A. Rusu, J.~Veness, M.~G. Bellemare,
  A.~Graves, M.~Riedmiller, A.~K. Fidjeland, G.~Ostrovski, S.~Petersen,
  C.~Beattie, A.~Sadik, I.~Antonoglou, H.~King, D.~Kumaran, D.~Wierstra,
  S.~Legg, and D.~Hassabis.
\newblock {Human-level control through deep reinforcement learning}.
\newblock \emph{Nature}, 518\penalty0 (7540):\penalty0 529--533, 2015.
\newblock ISSN 14764687.
\newblock \doi{10.1038/nature14236}.
\newblock URL \url{http://dx.doi.org/10.1038/nature14236}.

\bibitem[Poirazi et~al.(2003{\natexlab{a}})Poirazi, Brannon, and
  Mel]{Poirazi2003}
P.~Poirazi, T.~Brannon, and B.~W. Mel.
\newblock {Pyramidal neuron as two-layer neural network}.
\newblock \emph{Neuron}, 37\penalty0 (6):\penalty0 989--999,
  2003{\natexlab{a}}.
\newblock ISSN 08966273.
\newblock \doi{10.1016/S0896-6273(03)00149-1}.

\bibitem[Poirazi et~al.(2003{\natexlab{b}})Poirazi, Brannon, and
  Mel]{Poirazi2003a}
P.~Poirazi, T.~Brannon, and B.~W. Mel.
\newblock {Arithmetic of subthreshold synaptic summation in a model CA1
  pyramidal cell}.
\newblock \emph{Neuron}, 37\penalty0 (6):\penalty0 977--987,
  2003{\natexlab{b}}.
\newblock ISSN 08966273.
\newblock \doi{10.1016/S0896-6273(03)00148-X}.

\bibitem[Rall(1959)]{Rall1959}
W.~Rall.
\newblock {Physiological Properties of Dendrites}.
\newblock \emph{Annals of the New York Academy of Sciences}, 96\penalty0
  (4):\penalty0 1071--1092, 1959.

\bibitem[Schiller et~al.(2000)Schiller, Major, Koester, and
  Schiller]{Schiller2000}
J.~Schiller, G.~Major, H.~J. Koester, and Y.~Schiller.
\newblock {NMDA spikes in basal dendrites}.
\newblock \emph{Nature}, 1261\penalty0 (1997):\penalty0 285--289, 2000.
\newblock ISSN 1863-9135.
\newblock \doi{10.1127/1863-9135/2007/0169-0223}.

\bibitem[Segev(2006)]{segev2006}
I.~Segev.
\newblock {What do dendrites and their synapses tell the neuron?}
\newblock \emph{Journal of Neurophysiology}, 95\penalty0 (3):\penalty0
  1295--1297, 2006.
\newblock ISSN 0022-3077.
\newblock \doi{10.1152/classicessays.00039.2005}.
\newblock URL
  \url{http://jn.physiology.org/cgi/doi/10.1152/classicessays.00039.2005}.

\bibitem[Stuchlik(2014)]{Stuchlik2014}
A.~Stuchlik.
\newblock {Dynamic learning and memory, synaptic plasticity and neurogenesis:
  An update}.
\newblock \emph{Frontiers in Behavioral Neuroscience}, 8\penalty0
  (APR):\penalty0 1--6, 2014.
\newblock ISSN 16625153.
\newblock \doi{10.3389/fnbeh.2014.00106}.

\bibitem[Toni et~al.(1999)Toni, Buchs, Nikonenko, Bron, and Muller]{Toni1999}
N.~Toni, P.~Buchs, I.~Nikonenko, C.~R. Bron, and D.~Muller.
\newblock {LTP promotes formation of multiple spine synapses between a single
  axon terminal and a dendrite}.
\newblock \emph{Nature}, 402\penalty0 (November):\penalty0 421--425, 1999.

\bibitem[Travis et~al.(2005)Travis, Ford, and Jacobs]{travis2005regional}
K.~Travis, K.~Ford, and B.~Jacobs.
\newblock Regional dendritic variation in neonatal human cortex: a quantitative
  golgi study.
\newblock \emph{Developmental neuroscience}, 27\penalty0 (5):\penalty0
  277--287, 2005.

\bibitem[Ujfalussy et~al.(2015)Ujfalussy, Makara, Branco, and
  Lengyel]{Ujfalussy2015}
B.~B. Ujfalussy, J.~K. Makara, T.~Branco, and M.~Lengyel.
\newblock {Dendritic nonlinearities are tuned for efficient spike-based
  computations in cortical circuits}.
\newblock \emph{eLife}, 4\penalty0 (DECEMBER2015):\penalty0 1--51, 2015.
\newblock ISSN 2050084X.
\newblock \doi{10.7554/eLife.10056}.

\bibitem[Wilson et~al.(2016)Wilson, Whitney, Scholl, and
  Fitzpatrick]{Wilson2016}
D.~E. Wilson, D.~E. Whitney, B.~Scholl, and D.~Fitzpatrick.
\newblock {Orientation selectivity and the functional clustering of synaptic
  inputs in primary visual cortex}.
\newblock \emph{Nature Neuroscience}, 19\penalty0 (8):\penalty0 1003--1009,
  2016.
\newblock ISSN 15461726.
\newblock \doi{10.1038/nn.4323}.

\bibitem[Xiao et~al.(2017)Xiao, Rasul, and Vollgraf]{Xiao2017}
H.~Xiao, K.~Rasul, and R.~Vollgraf.
\newblock {Fashion-MNIST: a Novel Image Dataset for Benchmarking Machine
  Learning Algorithms}.
\newblock \emph{arXiv preprint}, pages 1--6, 2017.
\newblock URL \url{http://arxiv.org/abs/1708.07747}.

\bibitem[Zador and Pearlmutter(1996)]{Zador1996}
A.~M. Zador and B.~A. Pearlmutter.
\newblock {VC Dimension of an Integrate-and-Fire Neuron Model}.
\newblock \emph{Proceedings of the ninth annual conference on Computational
  learning theory}, pages 10--18, 1996.

\bibitem[Zador et~al.(1992)Zador, Claiborne, and Brown]{Zador1992}
A.~M. Zador, B.~J. Claiborne, and T.~H. Brown.
\newblock {Nonlinear pattern separation in single hippocampal neurons with
  active dendritic membrane}.
\newblock \emph{Advances in Neural Information Processing Systems}, pages
  51--58, 1992.

\end{thebibliography}

\setcounter{table}{0}
\renewcommand{\thetable}{S\arabic{table}}%
\setcounter{figure}{0}
\renewcommand{\thefigure}{S\arabic{figure}}%

\section{Supplementary Material}

\begin{figure}[ht]
    \centering
    \includegraphics[width=0.9\linewidth]{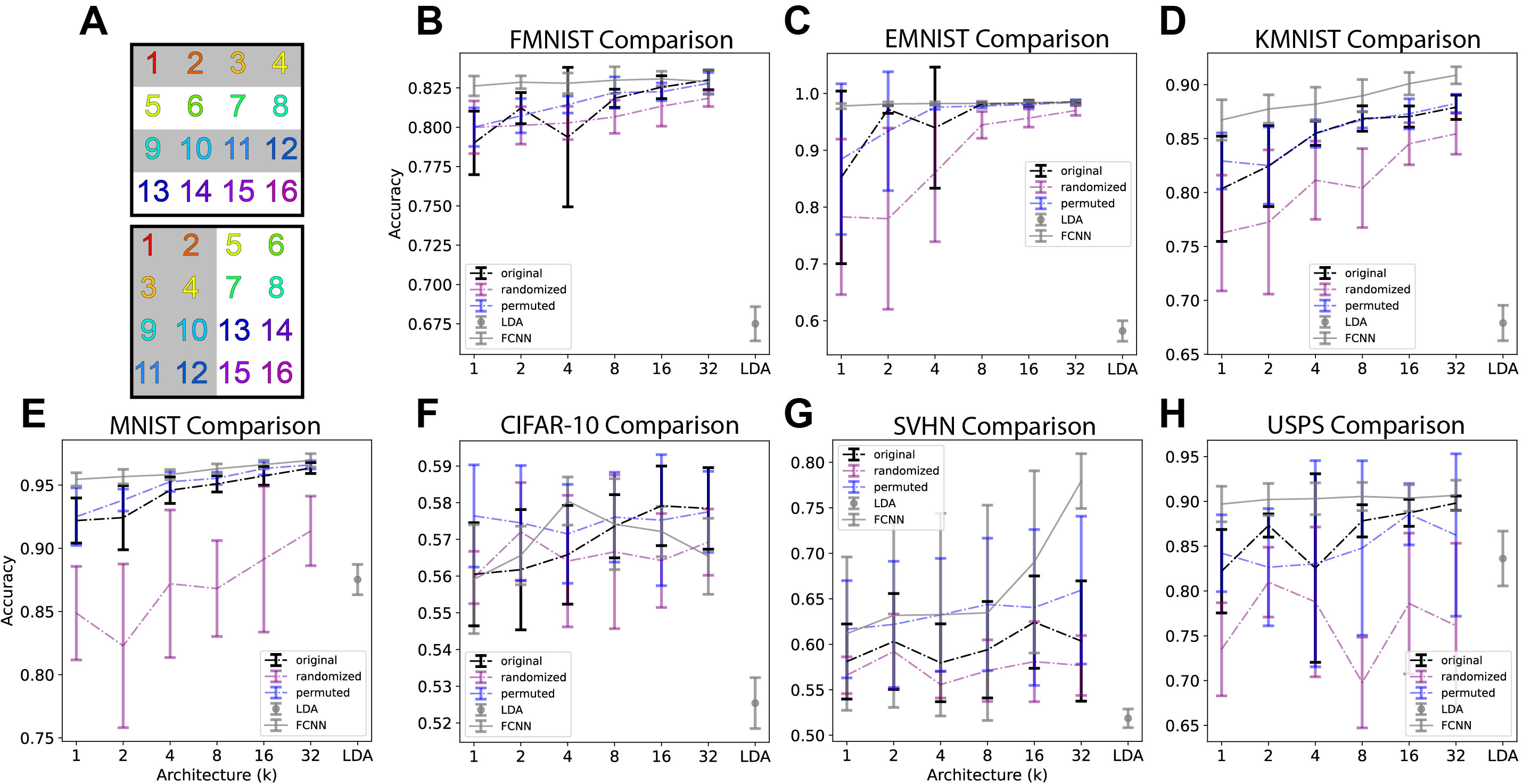}
    \caption{Permuted and randomized input trials. $k$-tree performance is compared to that of a lower bound, LDA, and an upper bound, FCNN. A: Example of original input setting (top) and permuted input setting (bottom). B-H: Original input is in black, permuted is in blue, randomized is in purple, FCNN and LDA are in gray. Randomized input tends to perform lower than original and permuted input. }
    \label{fig:inputs}
\end{figure}
\end{document}